\documentclass[11pt]{article}
\usepackage{graphicx}
\usepackage{array}

\newcommand{\BABARPubYear}    {02}

\newcommand{\BABARConfNumber} {026}
\newcommand{\SLACPubNumber} {9301}

\input babarsym
\newcommand{\etal}      {\mbox{\textsl{et al.}}\xspace}

 \def\xs         {\ensuremath{X_{s} }\xspace}
 \def\xsg        {\ensuremath{\xs \gamma}\xspace}
 \def\pizeta     {\ensuremath{\piz(\eta)}\xspace}

 \def\eg         {\ensuremath{E_{\gamma} }\xspace}
 
 \def\egcms      {\ensuremath{E^{*}_{\gamma}}\xspace}

 \def\mxs        {\ensuremath{m_{X_{s}} }\xspace}
 \def\mcut       {\ensuremath{m_{\mathrm{cutoff}} }\xspace}
 
 \def\ef         {\ensuremath{E_{f}^{*} }\xspace}
 \def\eb         {\ensuremath{E_{b}^{*} }\xspace}
 \def\rtwo       {\ensuremath{R_{2}^{*}}\xspace}
 \def\rtwoprime  {\ensuremath{R_{2}^{'}}\xspace}
 \def\rratio     {\ensuremath{R_{2}^{'}/R_{2}^{*} }\xspace}
 \def\pe         {\ensuremath{p_{e}^* }\xspace}
 \def\pmu        {\ensuremath{p_{\mu}^* }\xspace}
 \def\cose       {\ensuremath{\cos{\theta^{*}_{\gamma e}} }\xspace}
 \def\cosm       {\ensuremath{\cos{\theta^{*}_{\gamma \mu}} }\xspace}

 \def\mup        {\ensuremath{\mu_{\pi}}\xspace}
 \def\mb         {\ensuremath{m_{b} }\xspace}  
 \def\nc         {\ensuremath{\mathrm{n_C} }\xspace}
 \def\ns         {\ensuremath{\mathrm{n_S} }\xspace}
 \def\nb         {\ensuremath{\mathrm{n_B} }\xspace}
 \def\fOff       {\ensuremath{f_\mathrm{off} }\xspace}

 \def\bsg        {\ensuremath{b \to s \gamma}\xspace}
 \def\bdg        {\ensuremath{b \to d \gamma}\xspace}

 \def\bxsg       {\ensuremath{B \to X_{s} \gamma}\xspace}
 \def\bxdg       {\ensuremath{B \to X_{d} \gamma}\xspace}
 \def\bxg       {\ensuremath{B \to X \gamma}\xspace}

 \def\bkeg       {\ensuremath{\B \to \Kstar(892) \gamma }\xspace}

\long\def\inst#1{\par\nobreak\kern 4pt\nobreak
    {\it #1}\par\vskip 10pt plus 3pt minus 3pt}

\def\er #1 #2 { $#1 \pm #2$ }
\def\bra #1 #2 #3 #4 { $#1 ^{+#2} _{-#3} \pm #4 $ }
\def\pbxgresult {\ensuremath {[3.55 \pm 0.32(stat.) \pm 0.32(syst.)]\times 10^{-4}}\xspace}
\def\bsgresultuncor  {\ensuremath {[4.05 \pm 0.37(stat.) \pm 0.38(syst.) \pm ^{0.45}_{0.24}(model)]\times 10^{-4}}\xspace}
\def\bsgresultcor  {\ensuremath {[3.88 \pm 0.36(stat.) \pm 0.37(syst.) \pm ^{0.43}_{0.23}(model)]\times 10^{-4}}\xspace}
\def\bsgtheory  {\ensuremath {\BR(\bxsg)=3.45 \times 10^{-4}}\xspace}
\def\lumibb     {\ensuremath {(59.6 \pm 0.7) \times 10^6 \BB }\xspace}
\def\onlumipm   {\ensuremath {54.6 \pm 0.8  \invfb }\xspace}
\def\onlumi     {\ensuremath {54.6 \invfb }\xspace}
\def\offlumipm  {\ensuremath {6.40 \pm 0.10 \invfb }\xspace}

\def\sigrange   {\ensuremath {2.1 < \egcms < 2.7\gev}\xspace}

\begin{document}
{\pagestyle{empty}

%\begin{flushleft}
%BAD 464 Version 4.0
%\end{flushleft}

\begin{flushright}
\babar-CONF-\BABARPubYear/\BABARConfNumber \\
SLAC-PUB-\SLACPubNumber \\
%hep-ex/\LANLNumber \\
July  2002 \\
\end{flushright}

\par\vskip 5cm

% Title of the paper
\begin{center}
\Large \bf Determination of the Branching Fraction for Inclusive
Decays   $\bxsg$  
\end{center}
\bigskip

\begin{center}
\large The \babar\ Collaboration\\
\mbox{ }\\
July 24 2002
\end{center}
\bigskip \bigskip

% Abstract
\begin{center}
\large \bf Abstract
\end{center}

We present a preliminary determination of the inclusive branching fraction 
for the rare radiative penguin transition $\bxsg$. The measurement 
is based on a data sample of 60 million $\BB$ pairs collected between 
1999 and 2001 with the \babar\ detector at the \pep2 asymmetric-energy 
\epem \B Factory at SLAC.  We study events containing a high-energy 
photon from one \B (or \Bb) decay and a tagging primary lepton from 
the decay of the other \B meson. By this means, we are able to 
reduce a significant component of the background without  introduction of
model dependent uncertainties in the event selection efficiency. We determine 
the branching  fraction \BR(\bxsg)=\bsgresultcor, which is consistent with 
Standard Model predictions  and provides a constraint on possible new 
physics contributions to the electromagnetic penguin amplitude in $B$ decays.

\vfill
\begin{center}
Contributed to the 31$^{st}$ International Conference on High Energy Physics,\\ 
7/24---7/31/2002, Amsterdam, The Netherlands
\end{center}

\vspace{1.0cm}
\begin{center}
{\em Stanford Linear Accelerator Center, Stanford University, 
Stanford, CA 94309} \\ \vspace{0.1cm}\hrule\vspace{0.1cm}
Work supported in part by Department of Energy contract DE-AC03-76SF00515.
\end{center}

\newpage
} % end of pagestyle{empty}

% Input author list file
\begin{center}
\small

The \babar\ Collaboration,
\bigskip

%% author list as of 05-Jul-2002 (556 authors)
B.~Aubert,
D.~Boutigny,
J.-M.~Gaillard,
A.~Hicheur,
Y.~Karyotakis,
J.~P.~Lees,
P.~Robbe,
V.~Tisserand,
A.~Zghiche
\inst{Laboratoire de Physique des Particules, F-74941 Annecy-le-Vieux, France }
A.~Palano,
A.~Pompili
\inst{Universit\`a di Bari, Dipartimento di Fisica and INFN, I-70126 Bari, Italy }
% ...

\end{center}\newpage

% The body of the paper starts here
\section{Introduction}
\label{sec:Introduction}

The parton-level \bsg ``radiative penguin'' transition rate  
can be calculated at next-to-leading order to a  precision of 10\% in 
the Standard Model (SM)~\cite{bib:smtheory}. The presence of non-SM particles
in the virtual loop mediating this transition may lead to substantial
deviations from this predicted rate and has been the subject of numerous 
theoretical investigations~\cite{bib:nsmtheory}. The inclusive
\bxsg  rate is equal to the calculated parton-level \bsg rate according to 
quark-hadron duality~\cite{bib:smtheory}.
There have been several measurements of the branching fraction
for \bxsg to date~\cite{bib:otherexps}. 
The confinement of the \b quark in the \B meson
results in model dependent assumptions for the \eg spectrum, while the \xs
system fragments non-perturbatively into the detectable final state particles.
Thus, in order to access the parton model process, it is desirable to 
impose as few requirements as possible on the \xsg system, motivating a
fully-inclusive approach.  This, however, makes the measurement susceptible
to large backgrounds,
which must be suppressed without making such requirements.

In this paper we present a measurement of the branching fraction \BR(\bxsg)
in which a significant component of the background is removed by making
requirements on the other \B meson in the event, rather than on the
signal \bxsg decay. The remaining 
background is constrained by independent control samples. Although a 
requirement on \eg is still necessary, the accessible energy range is 
limited by the  statistics of the control samples, which scale directly 
with the signal size.  Hence this technique scales well with the 
increasing data sample anticipated at \babar.  This fully-inclusive technique
does not distinguish between decays of charged and neutral \B mesons, so
both are used.  For simplicity we refer to both \B and \Bb as
``\B mesons''.

\section{The \babar\ detector and dataset}
\label{sec:babar}

The data were collected with the \babar\ detector at the
 \pep2\ asymmetric-energy $\ep (3.1\gev ) $ -- $\en (9\gev ) $  
storage ring. A description of the \babar\ detector can be found in 
Ref.~\cite{bib:detector}. Charged particles
are detected and their momenta measured by a combination of
a silicon vertex tracker (SVT), consisting of
five double-sided layers, and a central drift chamber (DCH),
in a 1.5-T solenoidal field. We identify leptons and hadrons
with measurements from all detector systems,
including the energy loss (\dedx) in the DCH and SVT. Electrons
and photons are identified by a CsI electromagnetic calorimeter
(EMC). Muons are identified in the instrumented flux return (IFR).
A Cherenkov ring imaging detector (DIRC) covering the central region provides 
particle identification.

We use Monte Carlo simulations of the \babar\ detector 
based on GEANT 4.0~\cite{bib:geant}  to optimize our selection criteria, 
to determine signal efficiencies and to estimate part of the background 
component.  Events taken from random triggers are mixed with simulated events
to take beam backgrounds and some varying detector conditions into account.

The results  in this paper are based upon an
integrated luminosity of \onlumipm  of  data, corresponding to
\lumibb meson pairs recorded at the \FourS resonance (``on-resonance'') and
\offlumipm at $40\mev$ below this energy (``off-resonance''). The
off-resonance data, which constitute a fraction $\fOff = 10.5\%$ of the total
data set, are used for background estimation. The number of \BB\ meson pairs 
is determined from the excess hadronic events relative to muon pairs 
in on-resonance data compared to off-resonance data~\cite{bib:detector}.

\section{Analysis method}
\label{sec:Analysis}
\begin{figure}[htbp]
 \begin{center}
  \includegraphics[width=0.46\textwidth]{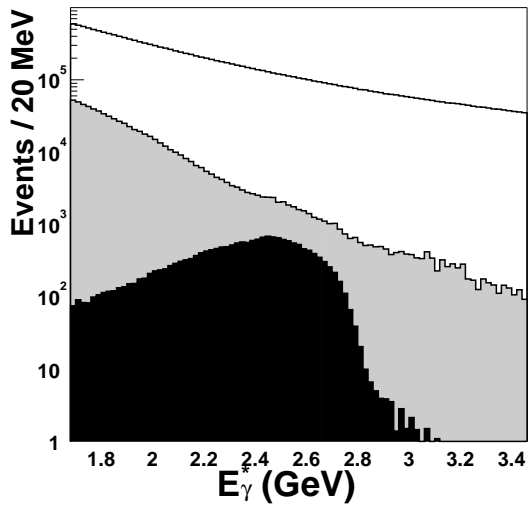}
  \caption {The energy distribution, in the \FourS center
   of mass, of simulated photon candidates, after ``photon quality'' cuts
   designed to reduce backgrounds from \piz{}s, $\eta$s and hadrons, but
   before other selection requirements.  Shown are
   \bxsg signal (dark shading), \BB background (grey shading) and
   continuum background (unshaded), all normalized to \onlumi.
   The high-end tail for \BB is mainly due to residual hadrons.}
  \label{fig:inclgammanocut}
 \end{center}
\end{figure}

The \bxsg decay signature is characterized by the high energy photon.
The background to the \bxsg\ signal consists of two components. First, it
can arise from other \B meson decays, in which the photon candidates are
predominantly from $\piz$ or $\eta$ (or rarely $\omega$) decays, 
or -- in a small fraction of cases -- from a hadronic interaction of
a neutron or $\KL$ in the electromagnetic calorimeter.  Second, it can arise
from continuum \qqbar or \tautau production, where $q$ can be a $u$,$d$,$s$ or 
$c$ quark, with the high-energy photon originating either from initial-state 
radiation (ISR) or from similar sources to those in \B decays.  
Figure~\ref{fig:inclgammanocut} demonstrates the scale of the background
rejection problem (based on Monte Carlo simulation).

The method for  extracting the signal from data is to subtract the
continuum background based on off-resonance data, and the \BB
contribution based on Monte Carlo predictions, where the latter are validated 
(and if necessary corrected) 
using our control samples.  A series of selection
requirements is made to greatly suppress backgrounds (primarily continuum)
before the subtraction.  
The selection criteria for this analysis are optimized for statistical 
precision by maximizing $\ns^{2}/(\ns+\nb+\nc/\fOff)$. The number 
of signal candidates  \ns  is estimated using a Monte Carlo simulation that 
incorporates the model of Kagan and Neubert~\cite{bib:smtheory} for the 
\eg spectrum and assumes 
a central theoretical value of \bsgtheory ~\cite{bib:smtheory}.
The \BB\ and continuum backgrounds, \nb and \nc, are estimated with Monte 
Carlo simulation. The signal and background expectations are normalized to 
\onlumi. The 
continuum is additionally weighted by 1/\fOff in the optimization to allow 
for the smaller size of the off-resonance data set.

Initially we require a high-energy photon candidate with 
$ 1.5 < \egcms < 3.5\gev $ in the \epem center-of-mass frame. 
(An asterisk denotes quantities computed in the \epem center-of-mass frame, 
as opposed to the laboratory frame.)
This is a loose selection that 
is tightened later to define the signal region and also control
``sideband'' regions.  A photon candidate is defined as a localized energy 
maximum~\cite{bib:detector} in the calorimeter acceptance  
$-0.74 <\cos{\theta} < 0.93$, where $\theta$ is the lab-frame polar angle 
relative to the \en beam direction.  It must be isolated by 40\cm 
at the inner surface of the EMC from any other photon 
candidate or track.  We veto candidates whose lateral energy profile 
is inconsistent with a single photon shower; these can arise from hadronic 
interactions in the calorimeter, or from \piz{}s in which the decay photons 
have merged into one shower (``merged \piz{}s''). In addition we veto 
photons from a $\piz(\eta)$ when the invariant mass of the 
combination with any other photon of energy greater than  50(250)\mev is 
within $\approx 2.7(2.2)$ sigma of the nominal $\piz(\eta)$ mass, 
$115(508) < M_{\gamma\gamma} < 155(588)\mevcc$.  However,  
we retain such vetoed photons as a control sample to validate the Monte
Carlo modeling of the \BB background. The photon candidates that fail the
lateral profile cut form the ``hadron anti-veto'' sample (it also includes
a small component of merged \piz{}s); while the 
photons that pass the lateral profile cut, but are rejected by the 
$\piz$ or $\eta$ veto, form the ``\pizeta anti-veto sample''. All three
independent samples, the signal and the two anti-veto samples, are 
required to pass all the subsequent selection criteria.  

  To suppress continuum backgrounds we require a ``lepton tag''. This
is a high-momentum electron ($\pe > 1.3\gevc$) or muon ($\pmu > 1.55\gevc$).
For \bxsg signal events the lepton arises from the semileptonic decay of the
other \B meson.
Leptons also occur in the continuum background, 
most notably from the semi-leptonic decays
of charm hadrons, but are produced significantly less frequently 
and with lower momentum than from a \B decay.
An electron candidate must have a ratio of calorimeter energy to track 
momentum, an EMC cluster shape,  a DCH \dedx\ and a DIRC Cherenkov angle 
(if available) consistent with an electron. A muon candidate must satisfy 
requirements on the measured and expected number of interaction lengths  
penetrated, the position match between the extrapolated DCH track and 
IFR hits, and the average and spread of the number of IFR hits per layer.  
The lepton candidates that 
do arise from continuum events are often pions misidentified as muons or,
less frequently, electrons. 
Fake leptons are usually not associated with an undetected neutrino, unlike
real leptons from a semi-leptonic \B decay.  
We therefore require that the missing energy in the event be greater 
than $1.2\gev$.
Further discrimination is obtained by considering the angle 
between the lepton and the high-energy photon.  For signal events, the
cosine of this angle,
\cose(\cosm),  has an approximately flat distribution,
except for a small peak at $\cosm=-1$ arising from cases
where the muon candidate and photon both originate from the signal
\B decay. (This can occur when a pion from the \xs decay fakes a muon.) In
contrast, the continuum background is strongly peaked at $\cose(\cosm)=-1$
as a consequence of the jet-like topology, with the peak at $\cose(\cosm)=+1$
suppressed by the photon isolation requirement. 
We require $\cose(\cosm)>-0.75(-0.70)$.
This also gives a small suppression of the \BB\ background for cases
where the high energy photon and lepton arise from the same \B decay.
Note that, since the tagging requirement is imposed  on the ``other''
\B rather than on the signal \B,
we are able to reject continuum background without imposing requirements 
on the signal decay. The signal efficiency of the combined lepton tag, missing
energy and $\cose(\cosm)$ requirements is 5\%, while the continuum
background is suppressed by a factor of approximately 1200.  
The momentum-dependent electron and muon efficiencies used for these and other
Monte Carlo results have been measured in data with a sample of radiative 
Bhabha events and samples of $\epem \to \mumu\gamma$ 
and $\epem \to \epem\mumu$ events, respectively.
  
Further continuum rejection is gained by imposing selection criteria
on the event topology. In the center-of-mass system \BB\ decays are
isotropic, while the inclusive \pizeta and ISR components of the 
continuum background  have a two and three-jet topology, respectively.
We compute the ratio of the second to the 
zeroth order Fox-Wolfram moment ~\cite{bib:foxandwolfram} in
both the center-of-mass frame, \rtwo, and the rest frame of the system 
recoiling against the high energy photon, \rtwoprime. The recoil frame
recovers the two-jet topology for ISR events. We require $\rtwo < 0.45$ and
$\rratio < 1$. We compute the energy sums of all particles (except the photon
candidate)  
whose momentum vectors lie within $0\degrees-30\degrees$ and 
$140\degrees-180\degrees$
cones about the photon direction, referred to as \ef and \eb, respectively.
We require $\ef < 1.1\gev$ and $ 1.6 < \eb < 3.6\gev $.

\begin{figure}[htbp]
 \begin{center}
  \begin{minipage}[b]{0.46\textwidth}
   \includegraphics[width=\textwidth,clip]
    {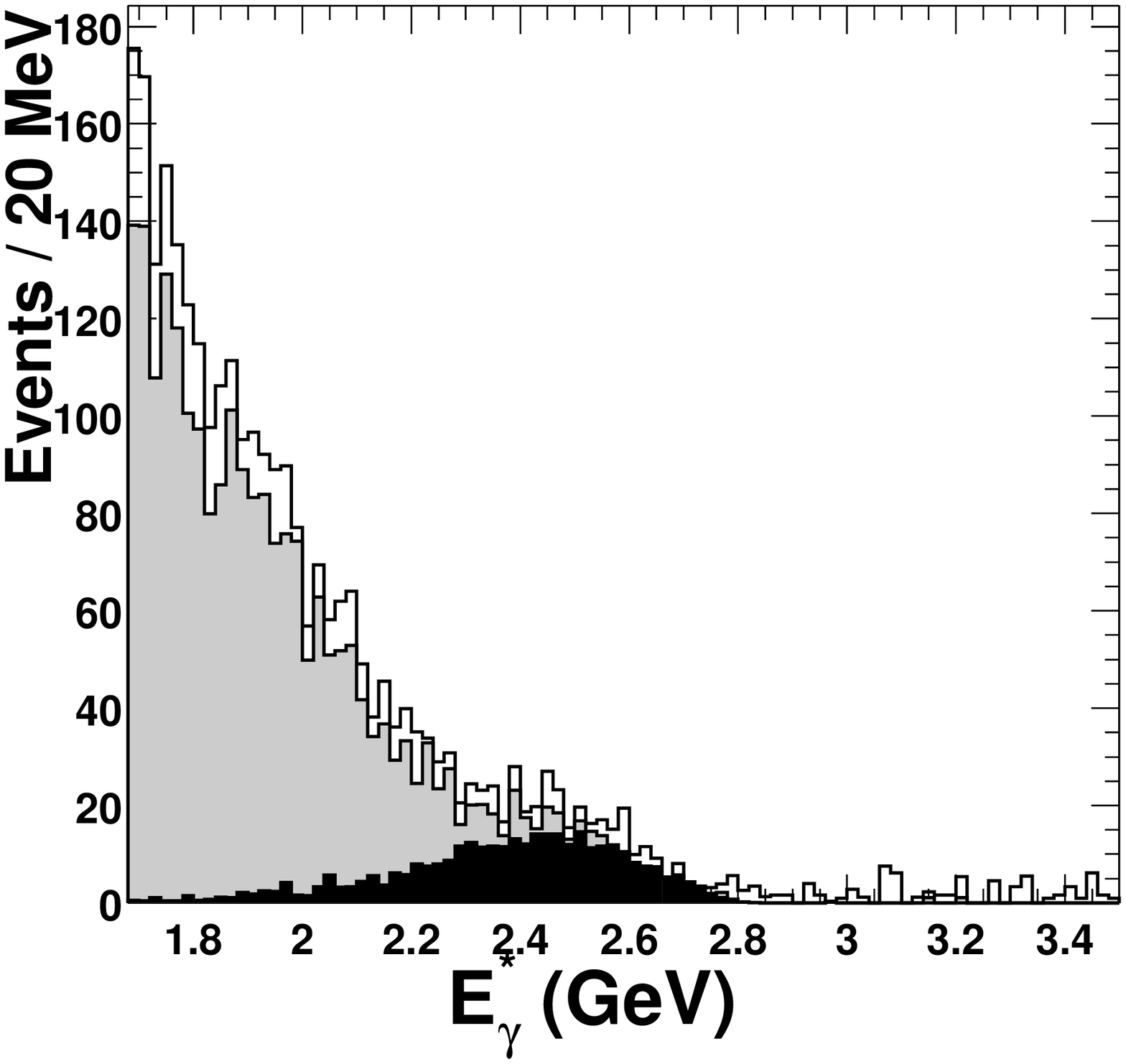} 
  \end{minipage}
  \begin{minipage}[b]{0.46\textwidth}
   \includegraphics[width=\textwidth,clip]
    {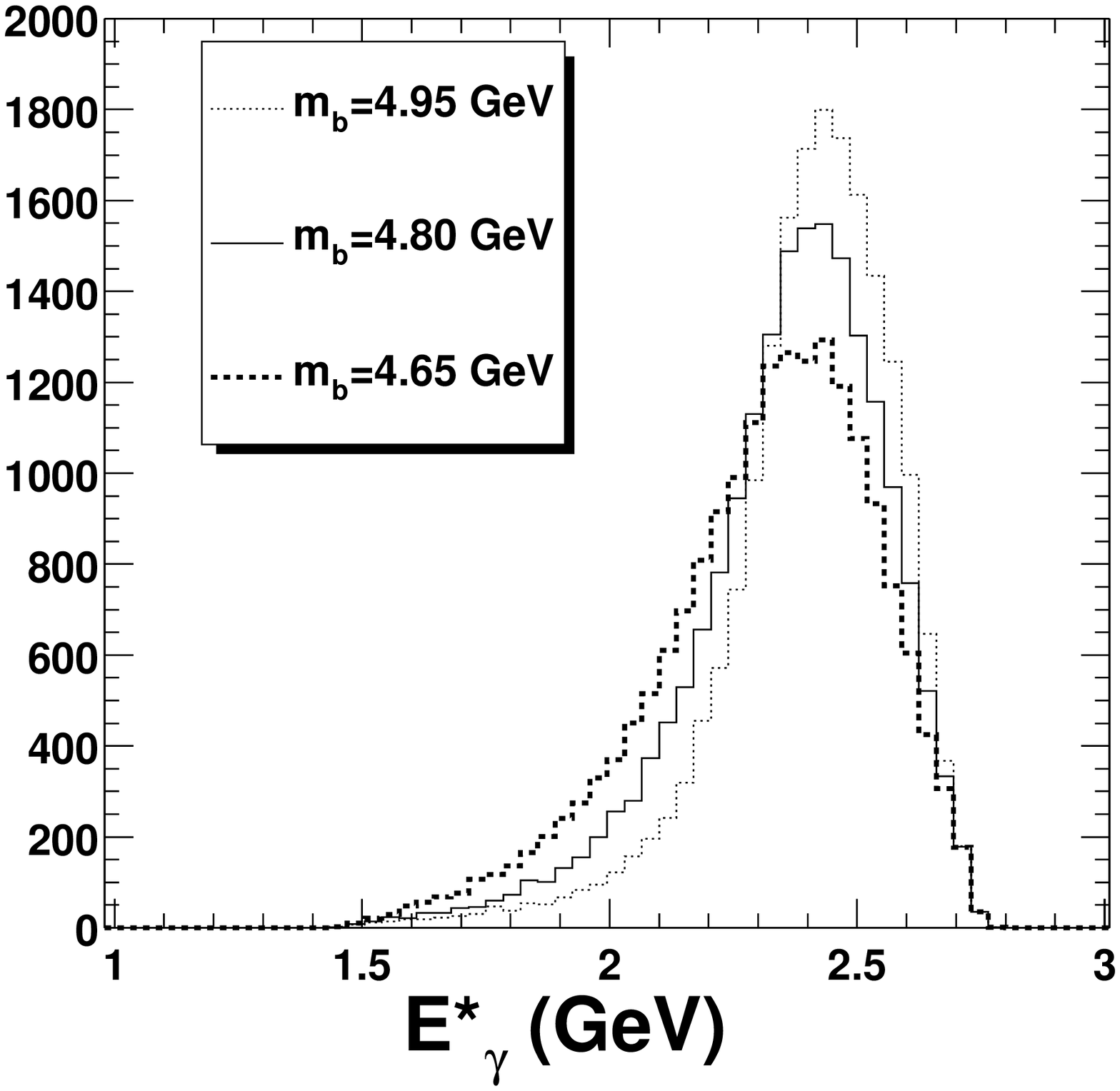}
  \end{minipage}
  \vspace{-0.15in}
  \caption{Left: The reconstructed \egcms distribution expected from Monte
   Carlo simulation after the selection criteria. The \bxsg signal 
   assuming \bsgtheory (dark shading), \BB background (grey shading) and 
   continuum background (unshaded) are normalized to \onlumi . 
   Right: The generated \egcms spectrum before cuts (arbitrary normalization)
   for different values of 
   the \b quark mass $\mb$, using the model of 
   Kagan and Neubert~\cite{bib:smtheory}.  Our signal region is defined
   for the corresponding reconstructed quantity as \sigrange.}
  \label{fig:mcexp}
 \end{center}
\end{figure}

Figure~\ref{fig:mcexp} shows the Monte Carlo prediction for the  \egcms 
distribution for signal, \BB background and continuum background after the
preceding selection criteria, and the expected \egcms signal distribution for 
different model parameters.  The signal region is defined to be
\sigrange, while the regions $1.7 < \egcms < 1.9$ and 
$2.9 < \egcms < 3.5~\gev$, which are dominated by \BB\ and continuum 
background,
respectively, serve  as  ``control'' regions for the estimation of these 
background components. The choice of the final $\egcms$ signal 
region, in particular the lower bound on \egcms, is a balance between the 
systematic uncertainty of the estimated \BB\ component, which is
the dominant systematic uncertainty of the measurement, and the 
model dependence. As the lower \egcms bound is reduced, the model
dependence decreases but the \BB\ background, which is estimated with
a Monte Carlo simulation,  rises sharply.

\section{Systematic studies}
\label{sec:Systematics}
\begin{table}[htbp]
\caption{The systematic error expressed as a percentage of $\BR(\bxsg)$. 
The total is the quadratic sum of all contributions.}
\begin{center}
\renewcommand{\arraystretch}{1.3}
\begin{tabular}{|l|c|} \hline 
Systematic Uncertainty        &  $\Delta\BR(\bxsg) / \BR (\%)$ \\ \hline \hline
\BB background estimation     &  $\pm 7.8 $        \\
Signal Monte Carlo statistics &  $\pm 2.6 $        \\
Photon efficiency             &  $\pm 2.5 $         \\
Lepton tag efficiency         &  $\pm 2.0 $         \\
Energy scale                  &  $\pm 1.0 $         \\  
Energy resolution             &  $\pm 1.0 $         \\ 
Photon distance requirement   &  $\pm 2.0 $         \\  
$\piz$/$\eta$ veto            &  $\pm 1.0 $         \\ 
\BB\ count                    &  $\pm 1.1 $         \\ \hline 
Total                         &  $\pm 9.3 $         \\ \hline
Signal model dependence       &  $+11.2,\ -\!5.9$    \\ \hline
\end{tabular} 
\end{center}
\label{tab:sys}
\end{table}

\begin{figure}[!htb]
\begin{center}
  \begin{minipage}[t]{0.46\textwidth}
   \vspace{0pt}
   \includegraphics[width=\textwidth,clip]
    {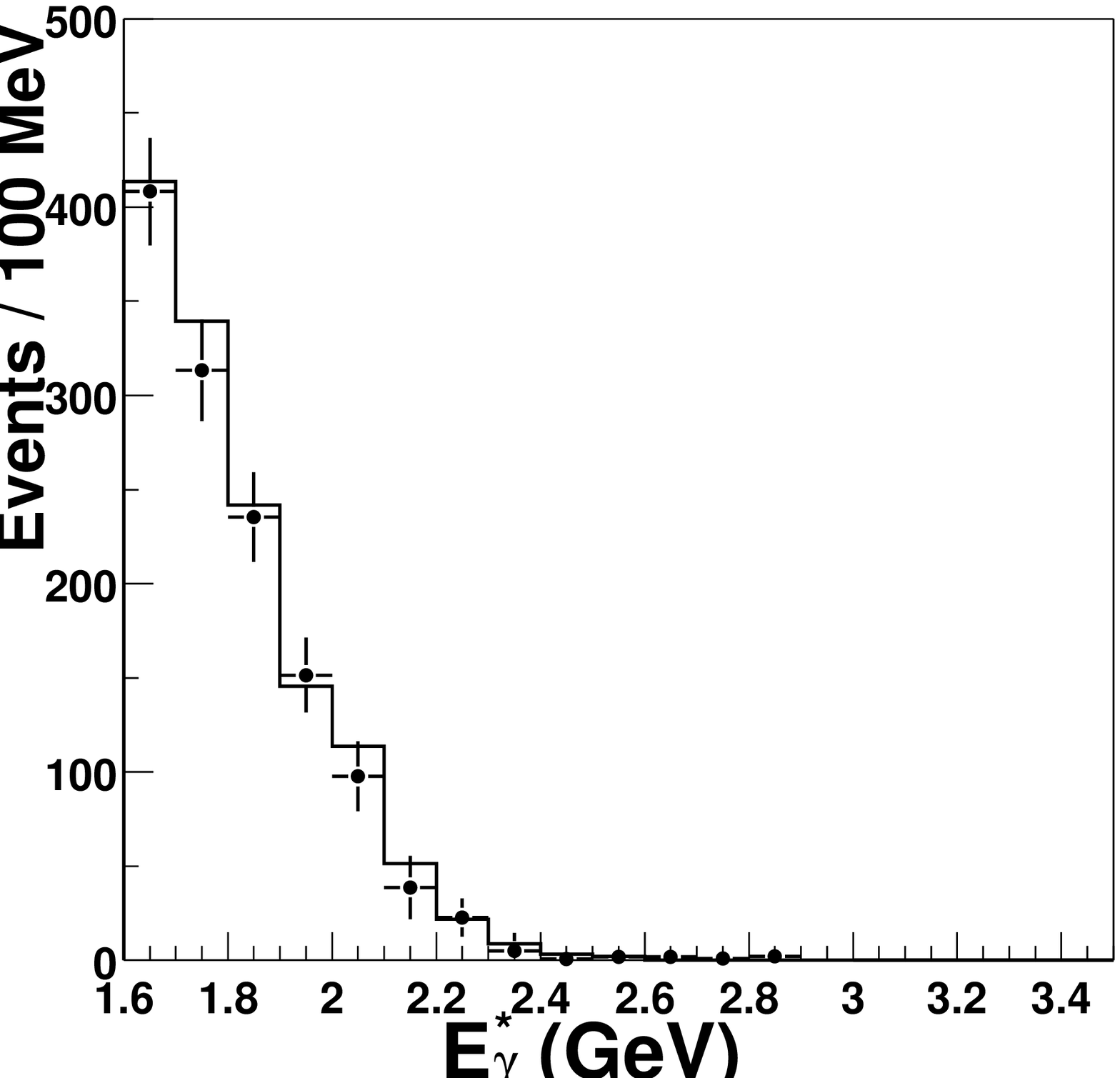}
  \end{minipage}
  \begin{minipage}[t]{0.46\textwidth}
   \vspace{11pt}
   \includegraphics[width=\textwidth,clip]
    {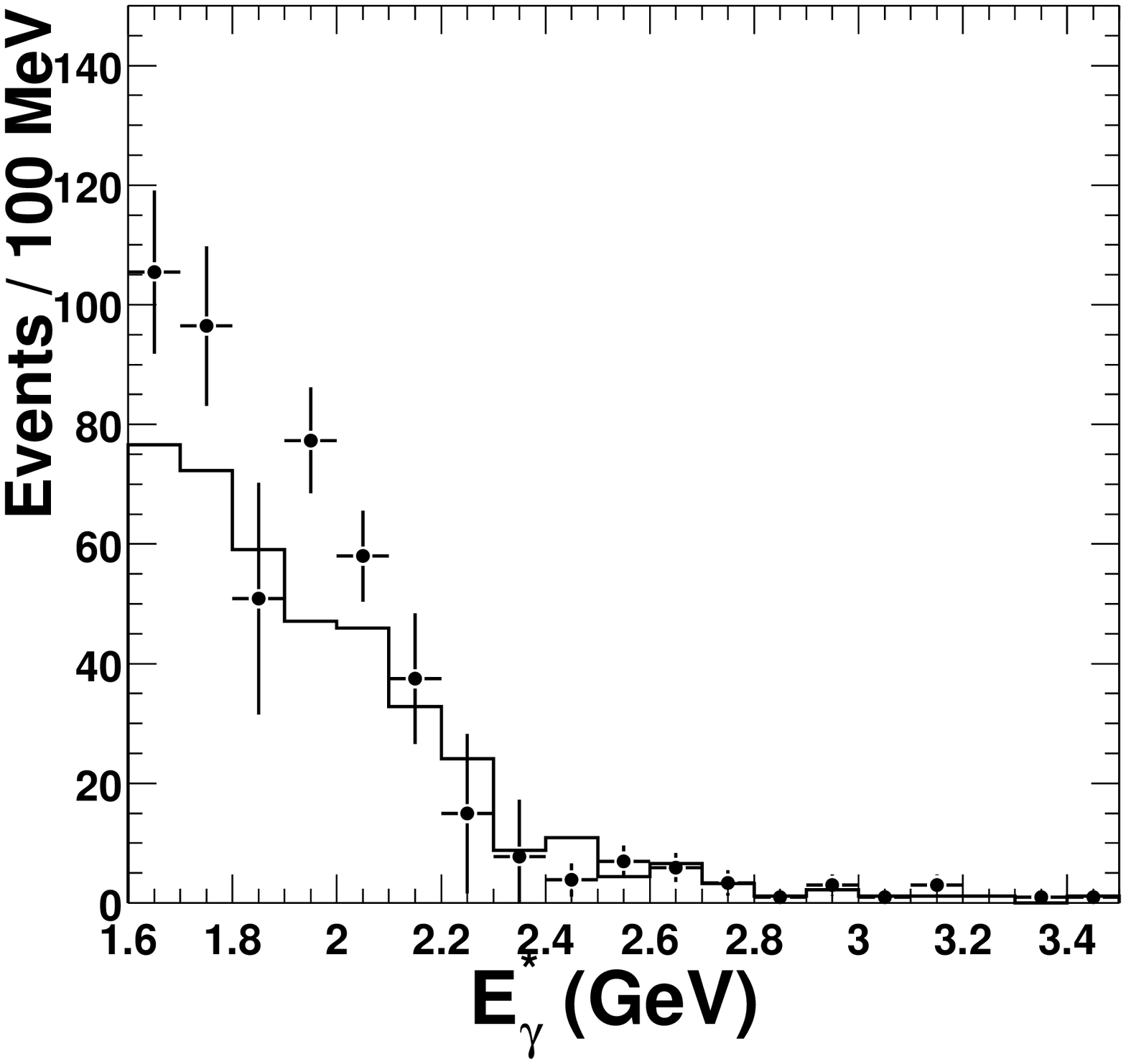}
  \end{minipage}
  \caption{Left: The \egcms distribution for  the $\piz$ and $\eta$  anti-veto 
   control sample for data (points) compared to the Monte Carlo prediction 
   (solid line). Right: Corresponding distribution for the hadron anti-veto 
   control sample.}
  \label{fig:pizantiveto}
 \end{center}
\end{figure}

In Table~\ref{tab:sys} we list the fractional systematic uncertainties
on the extracted branching fraction \BR(\bxsg). 
The dominant uncertainty is the modeling of the 
\BB background. To test the modeling of the \egcms spectrum
of the \BB background component, we compare the two anti-veto samples,
after the continuum contribution has been subtracted using the 
off-resonance data,  with generic \BB Monte Carlo event samples that have 
passed the same selection criteria
described above. Figure~\ref{fig:pizantiveto}  shows the comparison between 
Monte Carlo simulated events and data for the ``$\piz$ and $\eta$ anti-veto 
sample'' and the ``hadron anti-veto'' sample.  To take any differences 
into account, we separately fit the ratio of the measured data to Monte 
Carlo prediction for each control sample with a first-order polynomial,
and compute the average of the fit over the range \sigrange, weighted
by the \egcms spectrum expected for the corresponding background to
our signal sample.
We then correct the 
\egcms spectrum of the \BB\ Monte Carlo simulation used
to estimate the background in the signal region according to these
fit results, further weighted by the relative contributions of the 
$\piz$ and $\eta$ (91\%)
and hadronic and merged \piz components (9\%) to the background. 
To estimate the uncertainty
in this correction we find the variation incurred when using a zero-order 
polynomial fit rather than first-order, or moving the lower
bound of the fit region from 1.7 to 1.8\gev, or simply considering the
ratio of data to Monte Carlo events in the region \sigrange.  
This variation is summed in quadrature to
the statistical error of the first-order polynomial fit. The correction
factor for the total \BB background in the region \sigrange is 
$0.89 \pm 0.16$. The final uncertainty on the \BB\ background
estimate is computed by counting the expected number
of Monte Carlo events and then correcting as described, with the 
uncertainty being the sum in quadrature of the statistical error of the
Monte Carlo expectation and the uncertainty in the correction factor.
This results in a corrected expectation for \BB events in 
\sigrange of $222 \pm 43$.  (This is translated to the fractional entry in 
Table~\ref{tab:sys} by dividing the uncertainty by
the extracted signal from Section~\ref{sec:Physics}.)

The remaining systematic uncertainties are from the uncertainty in the
signal efficiency due to the photon selection and the lepton tag requirements,
plus a small contribution from \BB event counting.
The photon efficiency is measured
by comparing the ratio of events 
$N(\tau^{\pm} \to h^{\pm} \piz)/N( \tau^{\pm} \to h^{\pm} \piz \piz)$ 
to the previously measured branching fractions \cite{bib:pdg}. 
The photon isolation and $\piz/\eta$ veto efficiency
are dependent on the event multiplicity.  Their simulation is tested 
by ``embedding'' 
Monte Carlo-generated photons into both an exclusively reconstructed \B
meson data sample  and a generic \B meson Monte Carlo sample.
The photon-energy resolution is measured in data using $\piz$ and $\eta$ 
meson decays 
and a sample of virtual Compton scattering events. The energy scale 
uncertainty is estimated with a sample of $\eta$ meson decays with 
approximately
equal-energy photons; the deviation in the reconstructed $\eta$ 
mass from the nominal $\eta$ mass provides an estimate of the  uncertainty in 
the measured single photon energy.  Lepton tag systematics have been
estimated by coherently varying the measured efficiencies by $\pm 1\sigma$.
The total systematic uncertainty of 9.3\% is the quadratic sum of all
the contributions itemized in Table~\ref{tab:sys}.

We use an \egcms spectrum based on the  model of 
Kagan and Neubert~\cite{bib:smtheory} to determine the signal efficiency. 
In this model, the 
\eg spectrum,  which is dual to the mass spectrum of the \xs system through
the relation  $\eg = \frac{m^{2}_{B}-\mxs^{2}}{2m_{B}}$, where \eg is
the energy of the photon in the \B meson rest frame, 
has two components. The region $\mxs < \mcut$
is described by a relativistic Breit-Wigner for the \bkeg decay. The region 
$\mxs > \mcut$ is described by a spectrum parameterized in terms of the 
$b$-quark mass \mb.  The efficiency is sensitive to the value
of \mb because that determines how much of the spectrum falls outside
\sigrange (right plot in Figure~\ref{fig:mcexp}), but insensitive to
the fraction of $K^{*}(892)$ or to the \mcut value.  
Kagan and Neubert \cite{bib:smtheory} recommend varying \mb from 
4.65 to 4.95\gevcc to estimate model dependent variations.  We therefore
compute the efficiency of the signal using $\mb = 4.80 \pm 0.15\gevcc$.
We fix \mcut
to 1.1\gevcc, the value measured in a \babar\ semi-inclusive 
analysis of \bxsg \cite{bib:babarsemi}, and set the
$K^{*}(892)$ fraction equal to the integral of the discarded continuum
spectrum below \mcut for each \mb,
a procedure suggested by Kagan and Neubert, to find resulting shifts of 
+11.2\% and -5.9\% in \BR(\bxsg). To test the insensitivity of the result 
to the details of the relative contributions of \bkeg and \xs assumed in
the model, we explicitly vary \mcut by 
$\pm 100\mevcc$, and independently the fraction of \bkeg by a factor of 
two. We find negligible additional uncertainty on \BR(\bxsg).

\section{ Results}
\label{sec:Physics}

\begin{figure}[htbp]
 \begin{center}
   \includegraphics[width=0.5\textwidth,clip]
    {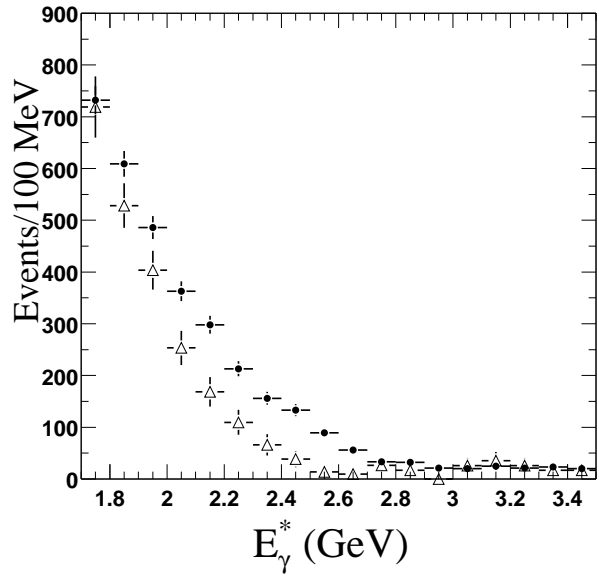}
  \vspace{-0.15in}
  \caption{The \egcms distribution of on-resonance data (solid points) compared to background expectation. All errors are statistical only (including, just
 for this figure, \BB background statistics).  Bin-by-bin
 systematic uncertainties and correlations have not yet been
 studied; the systematics quoted in the text apply \emph{only} to an integral 
 measurement from 2.1 to 2.7\gev.}
  \label{fig:data}
 \end{center}
\end{figure}

In order to reduce experimenter bias, the region $ 1.9 < \egcms < 2.9\gev $
was kept hidden until all selection criteria had been finalized and
backgrounds estimated.
Figure~\ref{fig:data} shows 
the \egcms spectrum for data compared to the total predicted background.
The control regions $ 1.7 < \egcms <  1.9\gev $ and 
$ 2.9 <  \egcms < 3.5\gev $ show no statistically significant excess of
data, as expected, but a signal is plainly visible in between.  
In the signal region, \sigrange, we find 543 net signal events.
Note that the selection procedure does not discriminate against the expected 
small component of \bdg, so these events are included in our total.
Combined with a selection efficiency \emph{within} the \egcms range 
of $1.28\pm 0.06\%$ and the
dataset size of \lumibb meson pairs, this yield corresponds to a  
measurement of the partial branching fraction for \sigrange 
of $\BR(\bxg)=\pbxgresult$.  For a Kagan and Neubert-based model with
$\mb = 4.80\gevcc$ (see Fig.~\ref{fig:mcexp}), the overall efficiency 
(including the effect of the \egcms cuts) is
$1.127 \pm 0.055\%$, resulting in $\BR(\bxg)=\bsgresultuncor$.
Finally, we subtract from this a contribution from $\bdg$.
In the Standard Model the theoretical expectation is  
$\BR(\bxdg)/\BR(\bxsg)=|V_{td}/V_{ts}|^{2}$. Assuming the signal efficiency 
for \bxdg to be equal to 
that for \bxsg and $|V_{td}/V_{ts}|=0.20 \pm 0.04$~\cite{bib:pdg} we then
scale the measured branching fraction by $0.96 \pm 0.02$ to give the
preliminary result:
\[
\BR(\bxsg)=\bsgresultcor. 
\]

\section{Summary}
\label{sec:Summary}

\begin{figure}[htbp]
 \begin{center}
   \includegraphics[width=0.7\textwidth,clip]
    {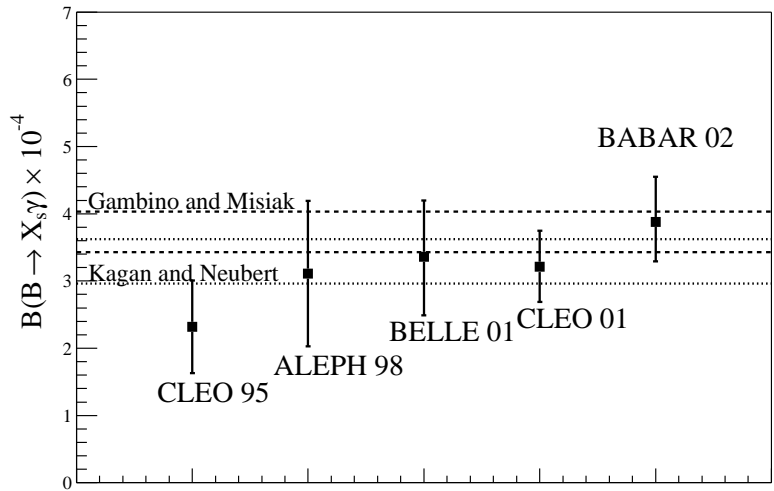}
  \vspace{-0.15in}
  \caption{The \babar\ measurement compared to previous 
experiments~\cite{bib:otherexps} and to 
theoretical predictions~\cite{bib:smtheory}.}
  \label{fig:bsgresult}
 \end{center}
\end{figure}

We find a preliminary branching fraction for \bxsg of  \bsgresultcor  
using an inclusive 
technique. Figure~\ref{fig:bsgresult} compares this result with
theoretical predictions and with previous measurements.  We do not see
any evidence for a departure from the next-to-leading order Standard
Model predictions. 
The systematic precision is limited by the size of the \BB\ 
background control samples, which scale in proportion to the signal sample. 
The systematic precision in turn limits the lower bound, $\egcms > 2.1\gev$, 
of the  signal region. As the larger datasets anticipated at the 
\B-factory become available, the systematic uncertainty in the \BB\ 
background will 
be reduced along with statistical uncertainties.  
Furthermore, this reduction  in the systematic
uncertainy may allow using a lower minimum-\egcms bound for the signal 
region, which will in turn lead to smaller model dependence.

\section{Acknowledgments}
\label{sec:Acknowledgments}

% Standard acknowledgments paragraph; must always be included.
We are grateful for the 
extraordinary contributions of our \pep2\ colleagues in
achieving the excellent luminosity and machine conditions
that have made this work possible.
The success of this project also relies critically on the 
expertise and dedication of the computing organizations that 
support \babar.
The collaborating institutions wish to thank 
SLAC for its support and the kind hospitality extended to them. 
This work is supported by the
US Department of Energy
and National Science Foundation, the
Natural Sciences and Engineering Research Council (Canada),
Institute of High Energy Physics (China), the
Commissariat \`a l'Energie Atomique and
Institut National de Physique Nucl\'eaire et de Physique des Particules
(France), the
Bundesministerium f\"ur Bildung und Forschung
(Germany), the
Istituto Nazionale di Fisica Nucleare (Italy),
the Research Council of Norway, the
Ministry of Science and Technology of the Russian Federation, and the
Particle Physics and Astronomy Research Council (United Kingdom). 
Individuals have received support from 
the A. P. Sloan Foundation, 
the Research Corporation,
and the Alexander von Humboldt Foundation.

%%%%%%%%%%%%%%%%%%--- References
%%%%%%%%%%%%%%%%%%%%%%%%%%%%%%%%%%%%%%%%%%%%%%%%%%%%%%%
\def\mpl #1 #2 #3 {Mod.~Phys.~Lett.~{\bf#1},\ #2 (#3)}
\def\npb  #1 #2 #3 {Nucl.~Phys.~B~{\bf#1},\ #2 (#3)}
\def\plb  #1 #2 #3 {Phys.~Lett.~B~{\bf#1},\ #2 (#3)}
\def\pr   #1 #2 #3 {Phys.~Rep.~{\bf#1},\ #2 (#3)}
\def\prd  #1 #2 #3 {Phys.~Rev.~D~{\bf#1},\ #2 (#3)}
\def\prl  #1 #2 #3 {Phys.~Rev.~Lett.~{\bf#1},\ #2 (#3)}
\def\RMP  #1 #2 #3 {Rev.~Mod.~Phys.~{\bf#1},\ #2 (#3)}
\def\zpc  #1 #2 #3 {Z.~Phys.~C~{\bf#1},\ #2 (#3)}
\def\nim  #1 #2 #3 {Nucl.~Instrum.~Methods~{\bf#1},\ #2 (#3)}
\def\nima  #1 #2 #3 {Nucl.~Instrum.~Methods~A.{\bf#1},\ #2 (#3)}
\def\epjc #1 #2 #3 {Euro.~Phys.~Jour~{\bf#1},\ #2 (#3)}
\def\rmp #1 #2 #3 {Rev.~Mod.~Phys~{\bf#1},\ #2 (#3)}
\def\npbps #1 #2 #3 {Nucl.~Phys.~B.~roc.~suppl~{\bf#1},\ #2 (#3)}
\def\progtp #1 #2 #3 {Prog.~Theo.~Phys~{\bf#1},\ #2 (#3)}

\end{document}